\documentclass[preprint,showpacs,preprintnumbers,superscriptaddress]{revtex4}
\usepackage{amssymb}
\usepackage{graphicx}
\usepackage{amsmath}
\usepackage{bm}
\usepackage{dcolumn}
\usepackage{amsfonts}
\begin{document}

\title{Shear and Bulk Viscosities of a Weakly Coupled Quark Gluon Plasma with Finite
Chemical Potential and Temperature---Leading-Log Results}
\author{Jiunn-Wei Chen}
\email{jwc@phys.ntu.edu.tw}
\affiliation{Department of Physics, Center for Theoretical Sciences, and Leung Center for
Cosmology and Particle Astrophysics, National Taiwan University, Taipei 10617, Taiwan}
\author{Yen-Fu Liu}
\email{linyenfu@gmail.com}
\affiliation{Department of Physics, Center for Theoretical Sciences, and Leung Center for
Cosmology and Particle Astrophysics, National Taiwan University, Taipei 10617, Taiwan}
\author{Yu-Kun Song}
\email{songyk@ustc.edu.cn}
\affiliation{Interdisciplinary Center for Theoretical Study and Department of Modern
Physics,University of Science and Technology of China, Anhui 230026, People's
Republic of China}
\affiliation{Key Laboratory of Quark and Lepton Physics (Central China Normal University),
Ministry of Education, People's Republic of China}
\author{Qun Wang}
\email{qunwang@ustc.edu.cn}
\affiliation{Interdisciplinary Center for Theoretical Study and Department of Modern
Physics,University of Science and Technology of China, Anhui 230026, People's
Republic of China}

\date{\today}

\begin{abstract}
We calculate the shear ($\eta$) and bulk ($\zeta$) viscosities of a weakly
coupled quark gluon plasma at the leading-log order with finite temperature
$T$ and quark chemical potential $\mu$. We find that the shear viscosity to
entropy density ratio $\eta/s$ increases monotonically with $\mu$ and
eventually scales as $\left(  \mu/T\right)  ^{2}$ at large $\mu$.\ In
contrary, $\zeta/s$ is insensitive to $\mu$. Both $\eta/s$ and $\zeta/s$ are
monotonically decreasing functions of the quark flavor number $N_{f}$ when
$N_{f}\geq 2$. This property is also observed in pion gas systems. Our
perturbative calculation suggests that QCD becomes the most perfect (i.e. with
the smallest $\eta/s$) at $\mu=0$ and $N_{f}=16$ (the maximum $N_{f}$ with
asymptotic freedom). It would be interesting to test whether the currently
smallest $\eta/s$ computed close to the phase transition with $\mu=0$ and
$N_{f}=0$ can be further reduced by increasing $N_{f}$.
\end{abstract}

\maketitle

\section{Introduction}

Viscosity, diffusivity, and conductivity are transport coefficients which
characterize the dynamics of long wavelength and low frequency fluctuations in a
medium. The quantity shear viscosity ($\eta$) per entropy density ($s$) has
attracted a lot of attention because of the intriguing conjecture that $\eta/s$
has a minimum bound of $1/4\pi$ for all systems \cite{Kovtun:2004de}. This
conjecture is inspired by the anti-de-Sitter space/conformal field theory
correspondence (AdS/CFT) \cite{Maldacena:1997re,Gubser:1998bc,Witten:1998qj}
which is rooted in string theory.  Surprisingly, the hot and dense matter
produced at RHIC \cite{Arsene:2004fa,Adcox:2004mh,Back:2004je,Adams:2005dq}
(for reviews, see e.g. Ref.\ \cite{Gyulassy:2004zy,Jacobs:2004qv,Heinz:2009xj,BraunMunzinger:2003zd})
just above the phase transition temperature ($T_{c}$) has $\eta/s=0.1\pm0.1$(theory)$\pm
0.08$(experiment) \cite{Luzum:2008cw}, a value close to the conjectured bound.
A robust limit of $1/(4\pi) \leq \eta /s \leq 2.5/(4\pi)$ at $T_c \leq T \leq 2T_c$ was recently extracted from
a VISHNU hybrid model
\cite{Song:2010mg} and a lattice computation of gluon plasma yields
$\eta/s=0.102(56)$ at temperature $T=1.24T_{c}$ \cite{Meyer:2007ic}.

The QCD transport coefficients have also been studied in other temperatures.
When $T\gg$ $T_{c}$, the $\eta$ of a weakly interacting quark gluon plasma is
inversely proportional to the scattering rate, $\eta\propto1/\Gamma
\propto1/\alpha_{s}^{2}\ln\alpha_{s}^{-1}$ \cite{Arnold:2000dr}, where
$\alpha_{s}$ is the strong coupling constant. The bulk viscosity $\zeta$ is
suppressed by an additional factor of $\left(  T_{\mu}^{\mu}\right)  ^{2}$,
arising from the response of the trace of the energy momentum tensor $\left(
T_{\mu}^{\mu}\right)  $\ to a uniform expansion. Thus, $\zeta$ vanishes when
the system is \textquotedblleft conformal\textquotedblright\ or scale
invariant. For a gluon plasma, the running of the coupling constant breaks the
scale invariance. Thus, $T_{\mu}^{\mu}\propto\beta\left(  \alpha_{s}\right)
\propto\alpha_{s}^{2}$, $\zeta\propto\alpha_{s}^{2}/\ln\alpha_{s}^{-1}$
\cite{Arnold:2006fz}. In the perturbative region, $\zeta/\eta\propto\alpha
_{s}^{4}\ll1$. When $T\ll$ $T_{c}$, the effective degrees of freedom are
pions. In the chiral limit ($u$ and $d$ quarks are massless), $\eta/s\propto
f_{\pi}^{4}/T^{4}$ \cite{Chen:2006iga} and $\zeta/s\propto T^{4}/f_{\pi}^{4}$
\cite{Chen:2007kx} where $f_{\pi}$ is the pion decay constant.
A compilation of perturbative QCD calculation of $\eta$ and $\zeta$
can be found, e.g., in Ref.\ \cite{Chen:2011km,Chen:2010xk}.
Most of these calculations are performed with finite $T$ but zero quark
chemical potential $\mu$.

The purpose of this work is to extend the previous perturbative QCD
calculation of $\eta$ and $\zeta$ to finite $\mu$ at the leading-log (LL)
order. At this order, we find $\eta\propto1/\alpha_{s}^{2}\ln\alpha_{s}^{-1}$ and
$\zeta\propto\alpha_{s}^{2}/\ln\alpha_{s}^{-1}$, same as in the limit of $\mu=0$,
which give parametrically the dominant contribution at asymptotically large
(compared with the $\Lambda_{QCD}$ scale where QCD becomes non-perturbative)
$T$ or $\mu$. The vacuum in our calculation has no spontaneous symmetry
breaking, thus, it cannot be applied to the color superconducting phase in the
$\mu/T \rightarrow \infty$ limit. In the context of finding the minimal $\eta/s$ and
hence the most \textquotedblleft perfect\textquotedblright\ fluid, we explore
whether $\eta/s$ can be further reduced by varying $\mu$ and the quark
flavor number $N_{f}$ in the hope that our perturbative calculation can shed
light on the non-perturbative region near $T_{c}$ where QCD is found to be the
most perfect matter ever produced in Nature.

\section{Effective Kinetic Theory}

$\eta$ and $\zeta$ can be calculated through the
linearized response function of thermal equilibrium states using the Kubo formula.
In the leading order (LO) expansion in the coupling constant,
the computation involves an infinite number of diagrams
\cite{Jeon:1994if,Jeon:1995zm,Carrington:1999bw,Wang:1999gv}.
However, it have been shown that the summation of the LO diagrams in a weakly coupled $\phi^{4}$
theory \cite{Jeon:1994if,Hidaka:2010gh} or in hot QED \cite{Gagnon:2007qt} is
equivalent to solving the linearized Boltzmann equation with
temperature-dependent particle masses and scattering amplitudes. This
conclusion is expected to hold in perturbative QCD as well.

The Boltzmann equation of a quark gluon plasma describes the evolution of the
color and spin averaged distribution function
$\tilde{f}_{p}^{a}(x)$ for particle $a$:
\begin{equation}
\frac{d\tilde{f}_{p}^{a}(x)}{dt} = \mathcal{C}_{a},
\end{equation}
where $\tilde{f}_{p}^{a}(x)$ is a function of space-time $x^{\mu}=(t,\mathbf{x})$
and momentum $p^{\mu}=(E_{p},\mathbf{p})$. For the LL calculation, we only
need to consider two particle scattering processes denoted as $ab\leftrightarrow cd$ with the
collision terms in the form
\begin{equation}
C_{ab\leftrightarrow cd}\equiv\int_{k_{1}k_{2}k_{3}}d\Gamma_{ab\text{
}\rightarrow cd}\left[  \tilde{f}_{k_{1}}^{a}\tilde{f}_{k_{2}}^{b} \tilde{F}_{p}^{c}  \tilde{F}_{k_{3}}^{d}
 -\tilde{F}_{k_{1}}^{a}\tilde{F}_{k_{2}}^{b}  \tilde{f}_{p}^{c}  \tilde{f}_{k_{3}}^{d}   \right]  .
\label{definition of C ab-cd}
\end{equation}
where $\tilde{F}^{g}=1+\tilde{f}^{g}$ and
$\tilde{F}^{q(\bar{q})}=1-\tilde{f}^{q(\bar{q})}$ and
\begin{equation}
d\Gamma_{ab\rightarrow cd}=\frac{1}{2E_{p}} \vert M_{ab
\rightarrow cd} \vert ^{2} \prod\limits_{i=1}^{3}\frac{d^3k_i
}{(2\pi)^3 2 E_{k_{i}}}(2\pi )^4 \delta ^{(4)} (k_{1}+k_{2}-k_{3}-p),
\label{gamma ab-cd}
\end{equation}
where $\vert M_{ab\rightarrow cd}\vert ^{2}$ is the matrix
element squared with all colors and helicities of the initial and final states summed over.
They are tabulated in Table \ref{amp} in the Appendix.
Then the collision term for a quark of flavor $a$ is
\begin{eqnarray}
N_{q}\mathcal{C}_{q_{a}}&=&\frac{1}{2}C_{q_{a}q_{a}\leftrightarrow q_{a}q_{a}}
+C_{q_{a}\bar{q}_{a}\leftrightarrow q_{a}\bar{q}_{a}}+\frac{1}{2}%
C_{gg\leftrightarrow q_{a}\bar{q}_{a}}+C_{q_{a}g\leftrightarrow q_{a}g}\nonumber\\
&&+\sum\limits_{b,b\neq a}(C_{q_{a}q_{b}\leftrightarrow q_{a}q_{b}%
}+C_{q_{a}\bar{q}_{b}\leftrightarrow q_{a}\bar{q}_{b}}+C_{q_{b}\bar{q}_{b}
\leftrightarrow q_{a}\bar{q}_{a}} ) ,
\end{eqnarray}
where $N_{q}=2\times3=6$ is the quark helicity and color degeneracy factor and the
factor $1/2$ is included when the initial state is formed by two
identical particles. Similarly,
\begin{equation}%
N_{g}\mathcal{C}_{g}=\frac{1}{2}C_{gg\leftrightarrow gg}
+\sum\limits_{a}(C_{gq_{a}\leftrightarrow gq_{a}}+C_{g\bar{q}_{a}
\leftrightarrow g\bar{q}_{a}}+C_{q_{a}\bar{q}_{a}\leftrightarrow gg} ),
\label{C22g}
\end{equation}
where $N_{g}=2\times8=16$ is the gluon helicity and color degeneracy factor. In
equilibrium, the distributions are denoted as $f^{q(\bar{q})}$ and $f^{g}$,
with
\begin{eqnarray}
f_{p}^{g}&=&\frac{1}{e^{u\cdot p/T}-1},
\label{distribution g} \\
f_{p}^{q(\bar{q})}&=&\frac{1}{e^{(u\cdot p\mp\mu )/T}+1},
\label{distribution q}
\end{eqnarray}
where $T$ is the temperature, $u$ is the fluid four velocity and $\mu$ is the
quark chemical potential. They are all space time dependent.

The thermal masses of gluon and quark or anti-quark for external states (the
asymptotic masses) can be computed via \cite%
{Arnold:2002zm,Mrowczynski:2000ed}
\begin{eqnarray}
m_{g}^{2} &=&\frac{2g^{2}}{d_{A}}\int \frac{d^{3}p}{(2\pi )^{3}2E_{p}}\Big[%
N_{g}C_{A}f_{p}^{g}+N_{f} N_{q}C_{F}(f_{p}^{q}+f_{p}^{%
\bar{q}})\Big],  \label{mg} \\
m_{q}^{2} &=&m_{\bar{q}}^{2}=2C_{F}g^{2}\int \frac{d^{3}p}{(2\pi )^{3}2E_{p}}%
(2f_{p}^{g}+f_{p}^{q}+f_{p}^{\bar{q}}),  \label{thermal mass g}
\end{eqnarray}%
where $d_{A}=8$, $C_{A}=3$, and $C_{F}=4/3$. This yields
\begin{eqnarray}
m_{g}^{2} &=&\frac{C_{A}}{6}g^{2}T^{2}+\frac{N_{f}C_{F}}{16}g^{2}(T^{2}+%
\frac{3}{\pi ^{2}}\mu ^{2}), \\
m_{q}^{2} &=&\frac{1}{4}C_{F}g^{2}\left( T^{2}+\frac{\mu ^{2}}{\pi ^{2}}%
\right) ,  \label{eqi thermal mass g}
\end{eqnarray}%
where we have set $E_{p}=|\mathbf{p}|$ in the integrals on the right hand
sides of Eqs.\ (\ref{mg}) and (\ref{thermal mass g}). The difference from
non-vanishing masses is of higher order.

\subsection{Linearized Boltzmann Equation}

To extract transport coefficients, it is sufficient to consider infinitesimal
perturbations away from equilibrium which have infinite wave lengths. Using
the Chapman-Enskog expansion we linearize the Boltzmann equation to the first
order in the derivative expansion in $x$. Thus, we only need the thermal equilibrium distributions $f^{g}$ and
$f^{q,\bar{q}}$ on the left-hand-side of the Boltzmann equation. We can also
make use of the zeroth-order energy-momentum conservation relation,
$\partial_{\mu}T^{(0)\mu\nu}=0$, to replace the time
derivatives $\partial T/\partial t$ and $\partial\mu/\partial t$ with spatial
gradients:
\begin{eqnarray}
\frac{\partial T}{\partial t}&=&-T\left( \frac{\partial P}{\partial\epsilon
}\right) _{n}\mathbf{\nabla\cdot u},\nonumber \\
\frac{\partial\mu}{\partial t}&=&-\left[  \mu\left(  \frac{\partial P}%
{\partial\epsilon}\right)  _{n}+\left(  \frac{\partial P}{\partial n}\right)
_{\epsilon}\right]  \mathbf{\nabla\cdot u},
\label{dT/dt}%
\end{eqnarray}
where $n\equiv n_{q}-n_{\bar{q}}$ is the baryon number density. We work in the
local rest frame of the fluid element where $u=(1,0,0,0)$ which implies
$\partial_{\mu}u^{0}=0$ after taking a derivative on $u^{\mu}(x)u_{\mu}(x)=1$.
For the right-hand-side of the Boltzmann equation, we expand the
distribution function of particle $a$ as a local equilibrium distribution plus
a correction
\begin{equation}
\tilde{f}^{a}\simeq f^{a}[1-\chi^{a}(1\pm f^{a})] ,
\label{deviation of f}
\end{equation}
and $\chi^{a}$ can be parametrized as
\begin{equation}
\chi^{a}(x,p) = \left[\frac{A^{a}(p)}{T}
\nabla_{i}u_{i}+\frac{B^{a}(p)}{T}\widehat{p}_{[i}
\widehat{p}_{j]}\nabla_{[i}u_{j]}\right]  ,
\label{khi}
\end{equation}
where $i,j=1,2,3$ and $[...]$ means the enclosed indices are made symmetric and traceless.

Applying Eq.(\ref{deviation of f}) to the right-hand-side of the Boltzmann
equation and equating it to left-hand-side, we get linear
equations for $B^{a}(p)  $ and $A^{a}(p)$. For $\tilde{f}^{g}$, we obtain
\begin{equation}
p_{[i}p_{j]}=\frac{E_{g}}{f^{g}F^{g}}\frac{1}{N_{g}}\Big[  \frac{1}%
{2}B_{gg\leftrightarrow gg}^{ij}+\sum\limits_{a=1}^{N_{f}}\left(
B_{gq_{a}\leftrightarrow gq_{a}}^{ij}+B_{g\overline{q}_{a}\leftrightarrow
g\overline{q}_{a}}^{ij}+B_{q_{a}\bar{q}_{a}\leftrightarrow gg}^{ij}\right)
\Big],
\label{linear equation for B from gluon}%
\end{equation}
where
\begin{equation}%
B_{ab\leftrightarrow cd}^{ij}(p)\equiv\int_{k_{1}k_{2}k_{3}}d\Gamma_{ab\text{
}\rightarrow cd}f_{k_{1}}^{a}f_{k_{2}}^{b}  F_{p}^{c} F_{k_{3}}^{d} \left[
-B_{ij}^{a}(k_{1})-B_{ij}^{b}(k_{2})+B_{ij}^{c}(p) +B_{ij}^{d}(k_{3}) \right]  .
\label{definition B ab-cd}%
\end{equation}
with $B_{ij}^{a}(p)=B^{a}(p)\widehat{p}_{[i}\widehat{p}_{j]}$
and we have suppressed the $p$ dependence in $B^a_{ij}$ and $B^a$
when there is no ambiguity. Similarly, for $\tilde{f}^{q}$, we obtain
\begin{eqnarray}
p_{[i}p_{j]}&=&\frac{E_{q}^{{}}}{f^{q}F^{q}}\frac{1}{N_{q}}\Big[  \frac{1}%
{2}B_{q_{1}q_{1}\leftrightarrow q_{1}q_{1}}^{ij}+B_{q_{1}\bar{q}%
_{1}\leftrightarrow q_{1}\bar{q}_{1}}^{ij}+\frac{1}{2}B_{gg\leftrightarrow
q_{1}\bar{q}_{1}}^{ij}+B_{q_{1}g\leftrightarrow q_{1}g}^{ij} \nonumber \\
&&+\sum\limits_{a=2}^{N_{f}}\left(  B_{q_{1}q_{a}\leftrightarrow
q_{1}q_{a}}^{ij}+B_{q_{1}\bar{q}_{a}\leftrightarrow q_{1}\bar{q}_{a}}
^{ij}+B_{q_{a}\bar{q}_{a}\leftrightarrow q_{1}\bar{q}_{1}}^{ij}\right)
\Big].
\label{linear equation for B from quark}%
\end{eqnarray}
The corresponding equation for $\tilde{f}^{\bar{q}}$ can be obtained from the
above equation by interchanging $q\leftrightarrow\bar{q}$. The linear
equations for $B^{a}\left(  p\right)  $ is relevant to the shear viscosity
computation. They can be written in a compact form
\begin{equation}
\left \vert S_{\eta}\right\rangle =\mathcal{C}_{\eta}\left\vert B\right\rangle ,
\label{linear equation for B}
\end{equation}
with $p_{[i}p_{j]}$ taken as the source $\left\vert S_{\eta}\right\rangle $
for the shear viscosity.

Similarly, the linear equations for $A^{a}(p)$ for quark and gluon are given by
\begin{eqnarray}
&&\frac{p^{2}}{3}-\left(  E_{g}^{^{2}}-T^{2}\frac{\partial m_{g}^{2}}{\partial T^{2}%
}-\mu^{2}\frac{\partial m_{g}^{2}}{\partial\mu^{2}}\right)  \left(  \frac{\partial
P}{\partial\epsilon}\right)  _{n}+E_{g}\left(  \frac{\partial E_{g}}%
{\partial\mu}-a_{g}\right)  \left(  \frac{\partial P}{\partial n}\right)
_{\epsilon}\nonumber \\
&&=\frac{E_{g}}{f^{g}F^{g}}\frac{1}{N_{g}}\left[  \frac{1}{2}
A_{gg\leftrightarrow gg}+\sum\limits_{a=1}^{N_{f}}\left( A_{gq_{a}
\leftrightarrow gq_{a}}+A_{g\overline{q}_{a}\leftrightarrow g\overline{q}_{a}
}+A_{q_{a}\bar{q}_{a}\leftrightarrow gg}\right) \right] ,
\label{linear equation for A from gluon}
\end{eqnarray}
and
\begin{eqnarray}
&&\frac{p^{2}}{3}-\left( E_{q}^{^{2}}-T^{2}\frac{\partial m_{q}^{2}}{\partial
T^{2}}-\mu ^{2}\frac{\partial m_{q}^{2}}{\partial \mu ^{2}}\right) \left( \frac{%
\partial P}{\partial \epsilon }\right) _{n}+E_{q}\left( \frac{\partial E_{q}%
}{\partial \mu }-a_{q}\right) \left( \frac{\partial P}{\partial n}\right)
_{\epsilon }  \notag \\
&=&\frac{E_{q}}{f^{q}F^{q}}\frac{1}{N_{q}}\left[ \frac{1}{2}%
A_{q_{1}q_{1}\leftrightarrow q_{1}q_{1}}+A_{q_{1}\bar{q}_{1}\leftrightarrow
q_{1}\bar{q}_{1}}+\frac{1}{2}A_{gg\leftrightarrow q_{1}\bar{q}%
_{1}}+A_{q_{1}g\leftrightarrow q_{1}g}\right.   \notag \\
&&\left. +\sum\limits_{a=2}^{N_{f}}\left( A_{q_{1}q_{a}\leftrightarrow
q_{1}q_{a}}+A_{q_{1}\bar{q}_{a}\leftrightarrow q_{1}\bar{q}_{a}}+A_{q_{a}%
\bar{q}_{a}\leftrightarrow q_{1}\bar{q}_{1}}\right) \right] ,
\label{linear equation for A from quark}
\end{eqnarray}%
where $a_{g}=0$, $a_{q/\bar{q}}=\pm 1$ and
\begin{equation}
A_{ab\leftrightarrow cd}(p)\equiv\int_{k_{1}k_{2}k_{3}}d\Gamma_{ab\text{
}\rightarrow cd}f_{k_{1}}^{a}f_{k_{2}}^{b} F_{k_{3}}^{c}  F_{p}^{d}  \left[
-A^{a}(k_{1})-A^{b}(k_{2})+A^{c}(k_{3})+A^{d}(p)  \right]  .
\label{definition A ab-cd}%
\end{equation}
These equations are relevant for the bulk viscosity computation. They can be
written compactly as
\begin{equation}
\left\vert S_{\zeta}\right\rangle = \mathcal{C}_{\zeta}\left\vert A\right\rangle .
\label{linear equation for A}%
\end{equation}
Using the expression of the thermal mass in Eq.\ (\ref{mg}),
we can evaluate the prefactor of $(\partial P/\partial \epsilon )_n$
in Eq.\ (\ref{linear equation for A from gluon}) as
\begin{eqnarray}%
E_{g}^{^{2}}-T^{2}\frac{\partial m_{g}^{2}}{\partial T^{2}}-\mu^{2}\frac{\partial
m_{g}^{2}}{\partial\mu^{2}}&=&\mathbf{p}^{2}-\beta(g^{2}) \left[
\frac{C_{A}}{6}T^{2}+\frac{N_{f}t_{F}}{6}\left( T^{2}+\frac{3}{\pi^{2}}%
\mu^{2}\right) \right] \nonumber \\
&\equiv & \mathbf{p}^{2}+\tilde{m}_{g}^{2},
\label{p^2+mg^2}
\end{eqnarray}
where $\alpha _{s}=g^{2}/4\pi $ and the QCD beta function for
 $g(\tilde{\mu}\equiv\sqrt{T^{2}+\mu^{2}})$ is
\begin{equation}
\beta (g^{2}) = \frac{\tilde{\mu}^{2}dg^{2}}{d\tilde{\mu}^{2}
}=\frac{g^{4}}{16\pi^{2}}\left( \frac{2N_{f}-33}{3}\right) .
\end{equation}
Similarly we obtain the prefactor of $(\partial P/\partial \epsilon )_n$
in Eq.\ (\ref{linear equation for A from quark}) as
\begin{eqnarray}
E_{q}^{^{2}}-T^{2}\frac{\partial m_{q}^{2}}{\partial T^{2}}-\mu^{2}\frac{\partial
m_{q}^{2}}{\partial\mu^{2}}&=&\mathbf{p}^{2}-\beta (g^{2})
\left[ \frac{1}{4}C_{F}\left(  T^{2}+\frac{\mu^{2}}{\pi^{2}}\right)  \right] \nonumber \\
&\equiv &\mathbf{p}^{2}+\tilde{m}_{q}^{2}.
\label{p^2+mq^2}%
\end{eqnarray}

As will be shown later, $(\partial P/\partial\epsilon )
_{n}-1/3\propto\beta (g^{2})$ and $(\partial P/\partial n )_{\epsilon}\propto\beta
(g^{2})$, thus the source of the bulk viscosity
$\left\vert S_{\zeta }\right\rangle \propto \beta (g^{2})$.
This means $\left\vert S_{\zeta}\right\rangle$ is,
as expected, proportional to the conformal symmetry breaking.

\subsection{Energy-Momentum Tensor and Quark Number Density}

The energy-momentum tensor of the kinetic theory can be written as \cite{Jeon:1994if}
(note the sign difference in metric with \cite{Jeon:1994if})
\begin{equation}
T^{\mu \nu }=\sum_{a}N_{a}\int \frac{d^{3}p}{(2\pi )^{3}}\frac{%
f^{a}(p,x)}{E_{a}}\left[ p^{\mu }p^{\nu }-\frac{1}{4}m_{a}^{2}(x)g^{\mu \nu }%
\right] ,  \label{energy momentum tensor}
\end{equation}%
where $a$ sums over the gluons and $N_f$ flavors of quarks and anti-quarks.
This equation expresses the total energy momentum tensor of the system as
the sum of individual quasiparticles. There is no $u^{\mu }u^{\nu }$ term on
the right hand side because energy momentum conservation cannot be satisfied
unless this term vanishes. In principle, one expects the form of Eq.\ (\ref%
{energy momentum tensor}) (and kinetic theory itself) will be no longer
valid at higher orders in the expansion of coupling constant, however,
Eq.(\ref{energy momentum tensor}) does reproduce all the
thermal dynamical quantities of QCD at $\mathcal{O}(\alpha _{s})$ correctly
\cite{Kapusta:1979fh}.

Expanding $T^{\mu\nu}$ to the first order of $\chi^{i}$, we have the fairly
simple expression
\begin{equation}%
\delta T^{\mu\nu}=-\sum_{a}N_{a}\int
\frac{d^{3}p}{(2\pi )^{3}E_{a}}f^{a}F^{a}\chi^{a}\left( p^{\mu}p^{\nu}
-u^{\mu}u^{\nu}T^{2}\frac{\partial E_a^{2}}{\partial T^{2}}
-u^{\mu}u^{\nu}\mu^{2}\frac{\partial E_a^2}{\partial\mu^{2}}\right) .
\label{energy momentum tensor diviation}%
\end{equation}
In deriving the above equation one needs to carefully
keep track of the implicit distribution function
dependence in $E_{a}$ through the thermal mass (\ref{thermal mass g}).
This expression can then be matched to hydrodynamics.

In hydrodynamics, $T^{\mu\nu}$ depends on $T$, $\mu$, and $u^{\mu}$. In the
local rest frame of the fluid element with $\mathbf{u}(x)=0$,
the most general form of $T^{ij}$ at the order of one
space-time derivative is (assuming parity and time reversal symmetry) can be decomposed into shear and bulk viscosity terms
\begin{equation}%
\delta T^{ij}=-2\eta\left(  \frac{\nabla_{i}u_{j}+\nabla_{j}u_{i}}{2}-\frac
{1}{3}\delta_{ij}\mathbf{\nabla}\cdot\mathbf{u}\right)  -\xi\delta
_{ij}\mathbf{\nabla}\cdot\mathbf{u}.
\label{definition of viscosity}%
\end{equation}
(Recall that $\partial T/\partial t$ and $\partial\mu/\partial t$ can be
replaced by $\mathbf{\nabla}\cdot\mathbf{u}$ as shown in Eq.(\ref{dT/dt}).)
Then using Eq.(\ref{khi}) for $\chi$ and comparing
Eqs.\ (\ref{energy momentum tensor diviation}) and
(\ref{definition of viscosity}), we obtain
\begin{equation}
\eta=\frac{1}{10T}\sum_{a}N_{a}\int
\frac{d^{3}p}{(2\pi)^{3}E_{a}}f^{a}F^{a}B_{jk}^{a}p_{[j}p_{k]},
\label{shear}
\end{equation}
and
\begin{equation}
\zeta=\frac{1}{T}\sum_{a}\int
\frac{d^{3}p}{(2\pi )^{3}E_{a}}\frac{p^{2}}{3}f^{a}F^{a}A^{a}.
\label{bulk}
\end{equation}
They can be written schematically as%
\begin{equation}
\eta=\langle B\vert S_{\eta}\rangle ,\ \zeta=\langle A\vert
S_{\zeta}^{\prime}\rangle .
\label{vis}
\end{equation}
Note that $\vert S_{\zeta}^{\prime}\rangle \neq \vert
S_{\zeta}\rangle $, but we will show $\zeta=\langle A\vert
S_{\zeta}^{\prime}\rangle =\langle A\vert S_{\zeta}\rangle $ later.

We also need the total quark number density
\begin{equation}
n=N_{f}N_{q}\int \frac{d^{3}p}{(2\pi )^{3}}(f^{q}-f^{\bar{q}}).
\label{quark number}
\end{equation}%
Expanding $n$ to the first order of $\chi ^{a}$, we have
\begin{equation}
\delta n=\sum_{a}N_{a}\int \frac{d^{3}p}{(2\pi )^{3}}f^{a}F^{a}\chi
^{a}\left( \frac{\partial E_{a}}{\partial \mu }-a_{a}\right) .
\label{delta quark number}
\end{equation}%
Eqs.\ (\ref{energy momentum tensor diviation}, \ref{delta quark number})
will be used in computation of bulk viscosity.

\subsection{Shear Viscosity}

It is well known that $\eta$\ should not be negative such that the second law
of thermodynamics (entropy cannot decrease in time) is satisfied.
This requirement is fulfilled by rewriting
Eqs.\ (\ref{linear equation for B},\ref{vis}) as
\begin{equation}
\eta=\langle B \vert \mathcal{C}_{\eta}\vert B \rangle
\label{vis1}
\end{equation}
and showing that $\eta$ is quadratic in $\left\vert B\right\rangle $ with a
positive prefactor such that it is bounded from below by zero. Indeed, these
conditions are satisfied in our expression:
\begin{eqnarray}
\eta &=&D_{gg\rightarrow gg}^{\eta}+\frac
{N_{f}(N_{f}-1)}{2}\left( 4D_{q_{a}q_{b}
\rightarrow q_{a}q_{b}}^{\eta}+4D_{\bar{q}_{a}\bar{q}_{b}
\rightarrow\bar{q}_{a}\bar{q}_{b}}^{\eta}+8D_{q_{a}\bar{q}_{b}\rightarrow q_{a}\bar{q}_{b}}^{\eta}%
+8D_{q_{a}\bar{q}_{a}\rightarrow q_{b}\bar{q}_{b}}^{\eta} \right)_{a\neq b} \nonumber\\
&& +N_{f}\left(  D_{qq\rightarrow qq}^{\eta}
+D_{\bar{q}\bar{q}\rightarrow\bar{q}\bar{q}}^{\eta}
+4D_{q\bar{q}\rightarrow q\bar{q}}^{\eta}
+4D_{gg\rightarrow q\bar{q}}^{\eta}+4D_{qg\rightarrow qg}^{\eta}
+4D_{\bar{q}g\rightarrow \bar{q}g}^{\eta}\right) ,
\label{shear'}
\end{eqnarray}
where there is no summation over $a$ and $b$ and we can just take $(a,b) = (1,2)$ and where
\begin{equation}
D_{ab\rightarrow cd}^{\eta}\equiv\frac{1}{80T}\int\frac{d^{3}p
}{( 2\pi )^{3}}d\Gamma_{ab\rightarrow cd}f_{k_{1}}^{a}f_{k_{2}}^{b} F_{k_{3}}^{c} F_{p}^{d}
\left[  B_{ij}^{a}(k_{1})+B_{ij}^{b}(k_{2})-B_{ij}^{c}(k_{3})-B_{ij}^{d}(p) \right]  ^{2}.
\label{definition Deta ab-cd}
\end{equation}
Once $\eta$ has the standard quadratic form in $\vert B\rangle $,
and $\eta=\langle B\vert \mathcal{C}_{\eta}\vert B\rangle
=\langle B\vert S_{\eta}\rangle $, we can use the standard algorithm to
systematically approach $\eta$ from below \cite{Chen:2011km}.

\subsection{Bulk Viscosity and the Landau-Lifshitz Condition}

For bulk viscosity, the collision kernel $\mathcal{C}_{\zeta}$ in Eq.\ (\ref{linear equation for A})
has two zero modes $A_{E}$ and $A_{n}$ which satisfy
\begin{equation}
\mathcal{C}_{\zeta}\vert A_{E}\rangle =\mathcal{C}_{\zeta}\vert A_{n}\rangle =0.
\label{definition of zero mode}
\end{equation}
$A_{E}$ arises from energy conservation
\begin{equation}
A_{E}^{a}(p)=E_{a},\ a=g,q,\bar{q},
\end{equation}
while $A_{n}$ arises from quark number conservation
\begin{equation}
A_{n}^{g}\left(  p\right)  =0,\ A_{n}^{q}\left(  p\right)  =1,
\ A_{n}^{\bar{q}}( p )  = - 1.
\label{zero mode An}
\end{equation}

We can use the Landau-Lifshitz condition
\begin{equation}
\delta T^{00}=0  \label{LS energy density constrain}
\end{equation}%
and
\begin{equation}
\delta n=0  \label{LS baryon number constrain}
\end{equation}%
to rewrite Eq.\ (\ref{bulk}) by adding linear combinations of $\delta T^{00}$
and $\delta n$:
\begin{eqnarray}
\zeta  &=&\sum_{a}\frac{N_{a}}{T}\int \frac{d^{3}p}{(2\pi )^{3}E_{a}}%
f^{a}F^{a}A^{a}\bigg[\frac{p^{2}}{3}-\left( \frac{\partial P}{\partial
\epsilon }\right) _{n}\left( E_{a}^{^{2}}-T^{2}\frac{\partial E_{a}^{2}}{%
\partial T^{2}}-\mu ^{2}\frac{\partial E_{a}^{2}}{\partial \mu ^{2}}\right)
\notag \\
&&+\left( \frac{\partial P}{\partial n}\right) _{\epsilon }E_{a}\left( \frac{%
\partial E_{a}}{\partial \mu }-a_{a}\right) \bigg]\equiv \langle A|S_{\zeta
}\rangle .  \label{bulk'}
\end{eqnarray}%
Substituting $\vert S_{\zeta}\rangle $ with
Eq.(\ref{linear equation for A}), $\zeta$ becomes quadratic in
$\left\vert A\right\rangle $ with a positive prefactor:%
\begin{eqnarray}
\zeta &=& \langle A\vert \mathcal{C}_{\zeta}\vert A\rangle \nonumber \\
&=& D_{gg\rightarrow gg}^{\zeta}+\frac{N_{f} (N_{f}-1)}{2}
\left(  4D_{q_{a}q_{b}\rightarrow
q_{a}q_{b}}^{\zeta}+4D_{\bar{q}_{a}\bar{q}_{b}\rightarrow\bar{q}_{a}\bar{q}_{b}}^{\zeta}
+8D_{q_{a}\bar{q}_{b}\rightarrow q_{a}\bar{q}_{b}}^{\zeta}
+8D_{q_{a}\bar{q}_{a}\rightarrow q_{b}\bar{q}_{b}}^{\zeta}\right)
_{a\neq b}\nonumber\\
&&+N_{f}\left(  D_{qq\rightarrow qq}^{\zeta}
+D_{\bar{q}\bar{q}\rightarrow\bar{q}\bar{q}}^{\zeta}
+4D_{q\bar{q}\rightarrow q\bar{q}}^{\zeta}
+4D_{gg\rightarrow q\bar{q}}^{\zeta}+4D_{qg\rightarrow qg}^{\zeta}
+4D_{\bar{q}g\rightarrow\bar{q}g}^{\zeta}\right)  ,
\label{bulk''}
\end{eqnarray}
where
\begin{equation}
D_{ab\rightarrow cd}^{\zeta}
\equiv
\frac{1}{8T}\int\frac{d^3p}{(2\pi)^{3}}
d\Gamma_{ab\rightarrow cd}f_{k_{1}}^{a}f_{k_{2}}^{b} F_{k_{3}}^{c} F_{p}^{d}
\left[  A^{a}(k_{1})+A^{b}(k_{2})-A^{c}(k_{3})-A^{d}( p ) \right]  ^{2}.
\label{definition Dkhi ab-cd}
\end{equation}
$(\partial P/\partial\epsilon )_{n}$ and $(\partial P/\partial n )_{\epsilon}$
in Eq.(\ref{bulk'}) can be obtained from thermodynamic relations
\begin{eqnarray}
\left(  \frac{\partial P}{\partial\epsilon}\right)  _{n}&=&\frac{sP_{,\mu,\mu
}-nP_{,\mu,T}}{C_{V}P_{,\mu,\mu}},\nonumber\\
\left(  \frac{\partial P}{\partial n}\right)  _{\epsilon}&=&\frac{nTP_{,T,T}
+(n\mu-sT)  P_{,\mu,T}-s\mu P_{,\mu,\mu}}{C_{V}P_{,\mu,\mu}},
\end{eqnarray}
where the pressure $P\left(  T,\mu\right)  $ can be read of from
Eq.\ (\ref{energy momentum tensor}) and where $P_{,X,Y}\equiv\partial
^{2}P/\partial X\partial Y$, $n=\partial P/\partial\mu$,
$s=\partial P/\partial T$, $\epsilon =-P+Ts+\mu n$
and $C_{V}=T (  P_{,T,T}-P_{,\mu ,T}^2/P_{,\mu,\mu})$.
We have also used
\begin{equation}
\left( \frac{\partial P}{\partial \epsilon }\right) _{n}=\left( \frac{\dfrac{%
\partial P}{\partial T}\delta T+\dfrac{\partial P}{\partial \mu }\delta \mu
}{\dfrac{\partial \epsilon }{\partial T}\delta T+\dfrac{\partial \epsilon }{%
\partial \mu }\delta \mu }\right) _{\delta n=\frac{\partial n}{\partial T}%
\delta T+\frac{\partial n}{\partial \mu }\delta \mu =0}.
\end{equation}
Then it can be shown that
\begin{equation}
\langle A_{E}\vert S_{\zeta}\rangle =0,\ \langle A_{n}\vert
S_{\zeta}\rangle =0.
\label{orthogonal between source and AE}
\end{equation}
Thus one can solve $\zeta = \langle A\vert \mathcal{C}_{\zeta}\vert A \rangle
=\langle A\vert S_{\zeta}\rangle$ for $\zeta$ using the standard algorithm to systematically
approach $\zeta$\ from below \cite{Chen:2011km}.

\subsection{Numerical Results}

As mentioned in the introduction, the LL result has $\eta\propto g^{-4}\ln^{-1}(  1/g)$
and $\zeta\propto g^{4}\ln^{-1}(1/g)$. When $\mu/T\ll1$, $\eta$, $\zeta$, and $s$ all scale as
$T^{3}$ from dimensional analysis. For $T/\mu\ll1$, $\eta$, $\zeta$, and $s$
should be even functions of $\mu$ because they should be invariant under
$\mu\rightarrow-\mu$, i.e. the exchange of quarks and antiquarks.

\begin{figure}[ptb]
\begin{center}
\includegraphics[height=2.6069in,width=4.1311in]{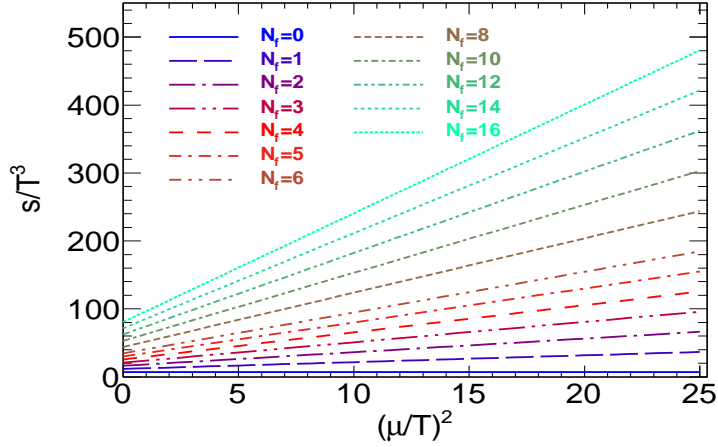}
\caption{(color online) $s/T^{3}$ as functions of $(\mu /T)^{2}$
for different $N_{f}$.}
\label{entropy density}
\end{center}
\end{figure}

We first show the result for the entropy density $s$. Only the leading order
$s$ for free particles is needed:
\begin{equation}
s=N_{g}s_{g}+N_{f}N_{q}\left( s_{q}+s_{\bar{q}}\right) ,  \label{s}
\end{equation}
where
\begin{eqnarray}
s_{g}&=&\int_p\left[\frac{\beta E_{p}}{e^{\beta E_{p}}-1}-\ln\left(1-e^{-\beta E_{p}}\right)\right] ,\\
s_{q,\bar{q}}&=&\int_p\left[\frac{\beta (E_{p}\mp\mu )}{e^{\beta(E_{p}\mp\mu)}+1}
+\ln\left(  1+e^{-\beta (  E_{p}\mp\mu)}\right)\right] .
\label{sg}%
\end{eqnarray}
In Fig.\ \ref{entropy density}, $s/T^{3}$ is shown as a function of $\left(
\mu/T\right)  ^{2}$. When $\mu/T\ll1$, $s$ scales as $T^{3}$ and when
$T/\mu\ll1$, $s$ scales as $\mu^{2}T$. This agrees with the expectation that
$s=0$ when $T=0$, and $s$ increases with $T$ for fixed $\mu$. The entropy density
$s$ also increases monotonically with the number of flavors $N_{f}$. We stop at
$N_{f}=16$, just before the asymptotic freedom of QCD is lost when
$N_{f}\geq 33/2$.

\begin{figure}[ptbptb]
\begin{center}
\includegraphics[height=2.7389in,width=4.1403in]
{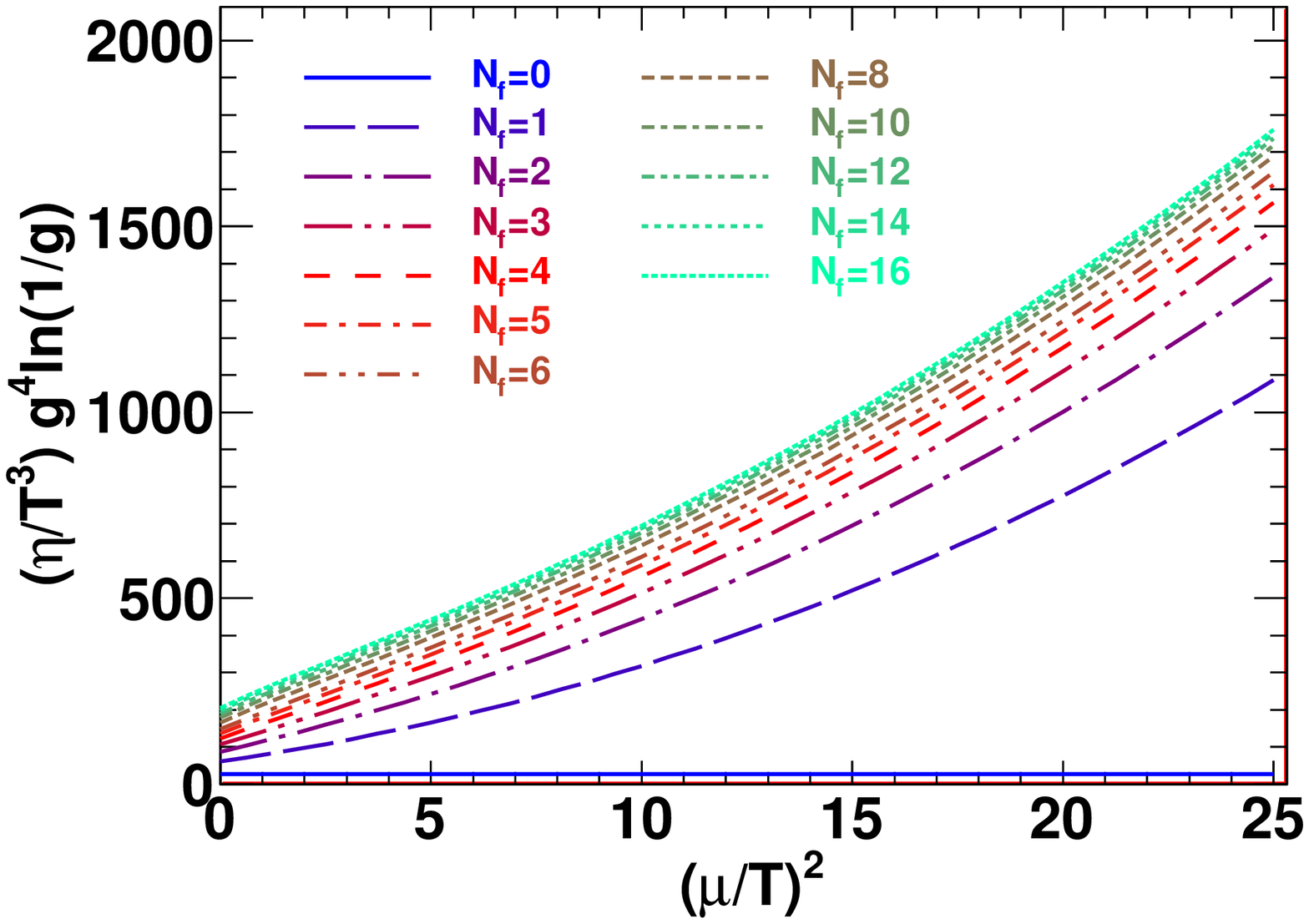}
\caption{(color online) $(\eta /T^{3})g^{4}\ln g^{-1}$ as
functions of $(\mu /T)^{2}$ for different $N_{f}$. }
\label{eta with musquare}
\end{center}
\end{figure}

In Fig. \ref{eta with musquare}, we show the normalized shear viscosity
$\tilde{\eta}\equiv (\eta/T^{3})g^{4}\ln g^{-1}$ as functions
of $(\mu/T)^{2}$ for different $N_{f}$. When $\mu/T\rightarrow0$,
$\eta$ scales as $T^{3}$ from dimensional analysis. When $T/\mu\rightarrow0$,
$\eta$ cannot scale as $\mu^{3}$ because it is an even function of $\mu$.
Instead, $\eta$ scales as $\mu^{4}/T$. Technically, this is because
$f^{q}F^{q}\propto\delta\left(  \left(  E-\mu\right)  /T\right)
=T\delta\left(  E-\mu\right)  $ as $T/\mu\rightarrow0$ while the anti-quark
and gluon contributions vanish. Thus, Eqs. (\ref{shear}) and (\ref{shear'}) are
solved with $B_{jk}\propto1/T$ and $\eta\propto1/T.$ Physically, this $1/T$
behavior emerges because $\eta$ is inversely proportional to the collision
rate which vanishes at $T=0$. Also, $\eta$ is monotonically increasing with
$N_{f}$ because the averaged coupling between gluons is stronger than those
with quarks involved. Thus, the effective collision rate is smaller with
higher $\mu$ and higher $N_{f}$.

Around $\mu=0$, we make a Taylor expansion $\tilde{\eta}=a_{\eta}+b_{\eta}(
\mu/T )^{2}+\cdots$, where $a_{\eta}$ is $\tilde{\eta}$ at zero quark chemical potential.
The values of $a_{\eta}$ and $b_{\eta}$ for various
$N_{f}$ is tabulated in Table \ref{power expansion coefficients eta}.
Our $a_{\eta}$ is identical to AMY's to at least the second decimal place for all $N_{f}$
computed in Ref.\ \cite{Arnold:2000dr}. The agreement is better than $1\%$.

\begin{figure}[ptb]
\begin{center}
\includegraphics[height=2.7464in,width=4.1494in]
{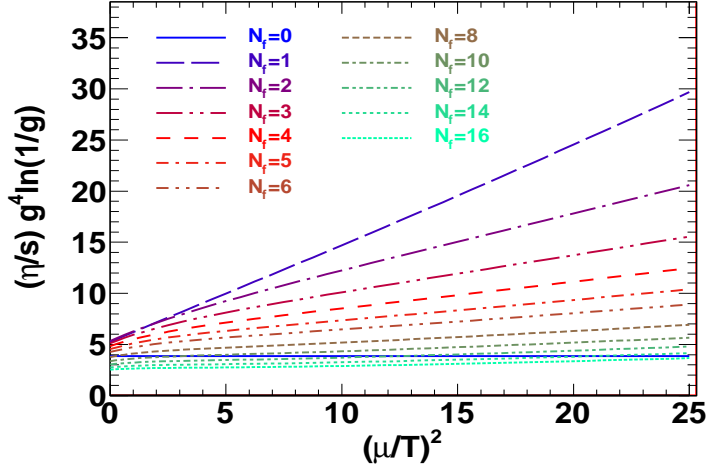}
\caption{(color online) $(\eta/s)g^{4}\ln g^{-1}$ as
functions of $(\mu/T)^{2}$ for different $N_{f}$.}
\label{eta over s with musquare}
\end{center}
\end{figure}

In Fig.\ \ref{eta over s with musquare}, $(\eta/s)g^{4}\ln g^{-1}$
is shown as functions of $(\mu/T)^{2}$. For a given coupling,
the LL value of $\eta/s$ is the smallest at $\mu =0$, i.e. the fluid is the most perfect
at $\mu=0$. This perturbative QCD result at high $T$ is consistent with the
observation of Ref.\ \cite{Chen:2007xe} in the hadronic phase at low $T$.
Thus, we speculate that this property might also be true near the phase transition
temperature $T_c$ such that QCD has its local minimum, perhaps its absolute minimum
as well, at $T_c$ with zero quark chemical potential.

If the coupling $g$ is held fixed, our $\eta/s$ is monotonically decreasing with $N_f$ for
$N_{f}\geq 2$ but not for $N_{f}=0$ and $1$ (there is a crossing between the $\eta/s$ of $N_{f}=1$ and
$2$  at $\mu^{2}/T^{2}\simeq1.8$). This pattern looks random, but
interestingly, it is qualitatively consistent with the pion gas result of
Ref.\ \cite{Chen:2006iga} which has $\eta/s\propto f_{\pi}^{4}/N_{f}^{2}T^{4}\propto
N_{c}^{2}/N_{f}^{2}$ (we have used $f_{\pi}^{2}\propto
N_{c}$). Again, this suggests that there is a natural connection
between $\eta/s$ above and below the phase transition. This pion gas
analogy can also explain why $N_{f}=0$ and $1$ are special---there is no
pion in these two cases.

We also observe that when $N_{f}\geq8$, the fluid can be more perfect than
that of $N_{f}=0$. It would be interesting to test this in lattice QCD to find
a more perfect fluid than the currently evaluated $N_{f}=0$ case. It would also be
interesting to investigate how the above qualitative $N_{f}$ scaling changes
due to the possible infrared fixed point for $N_{f}\gtrsim12$ where chiral
symmetry is not supposed to be broken and hence no pions exist anymore (see, e.g. \cite{Lin:2012iw}
and references therein).

\begin{figure}[ptbptb]
\begin{center}
\includegraphics[height=2.8202in,width=4.1669in]
{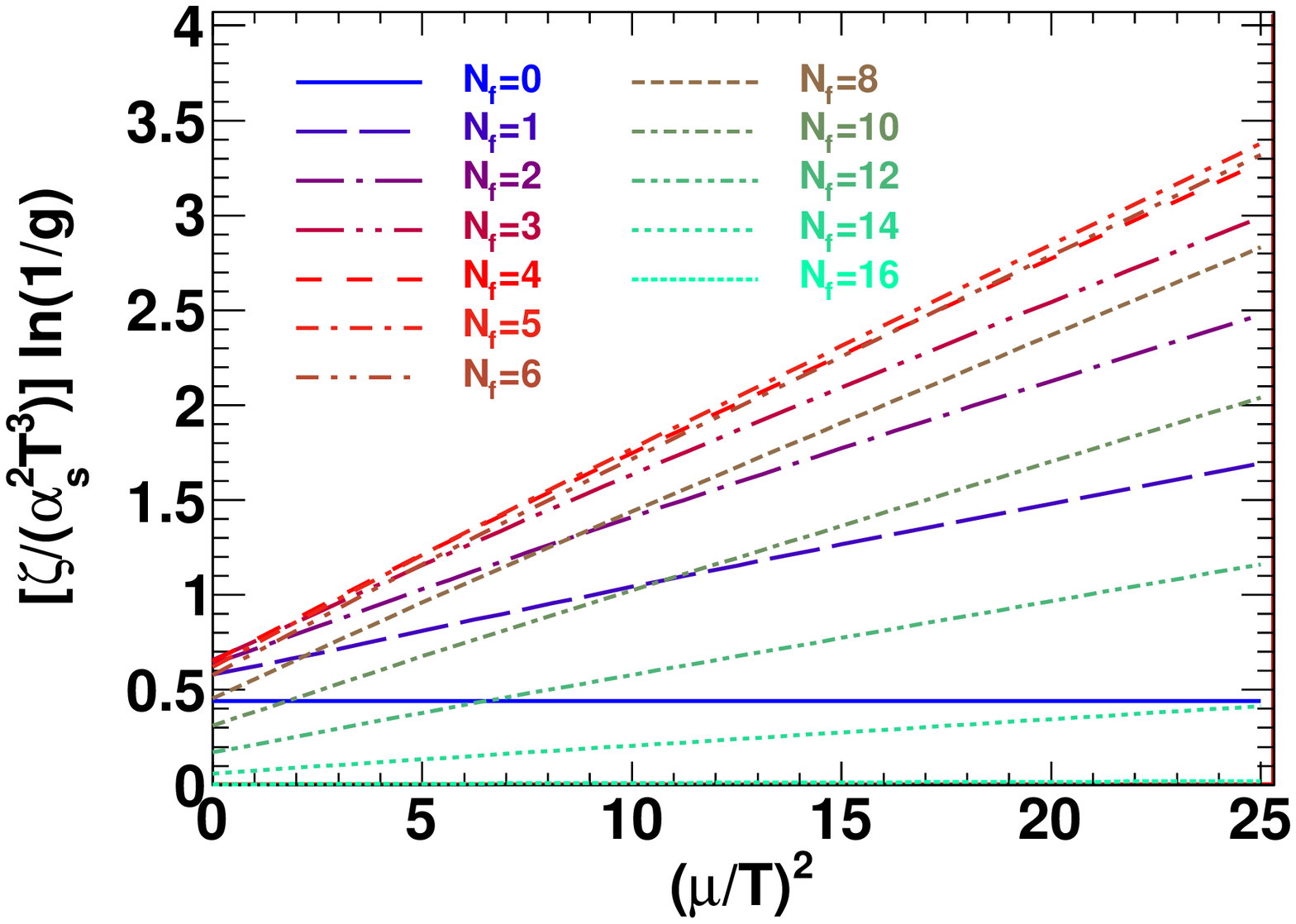}
\caption{(color online) $(\zeta/\alpha_{s}^{2}T^{3})\ln g^{-1}$
as functions of $(\mu/T)^{2}$ for different $N_{f}$.}
\label{bulk viscosity with musquare}
\end{center}
\end{figure}

\begin{figure}[ptbptbptb]
\begin{center}
\includegraphics[height=2.8028in,width=4.215in]
{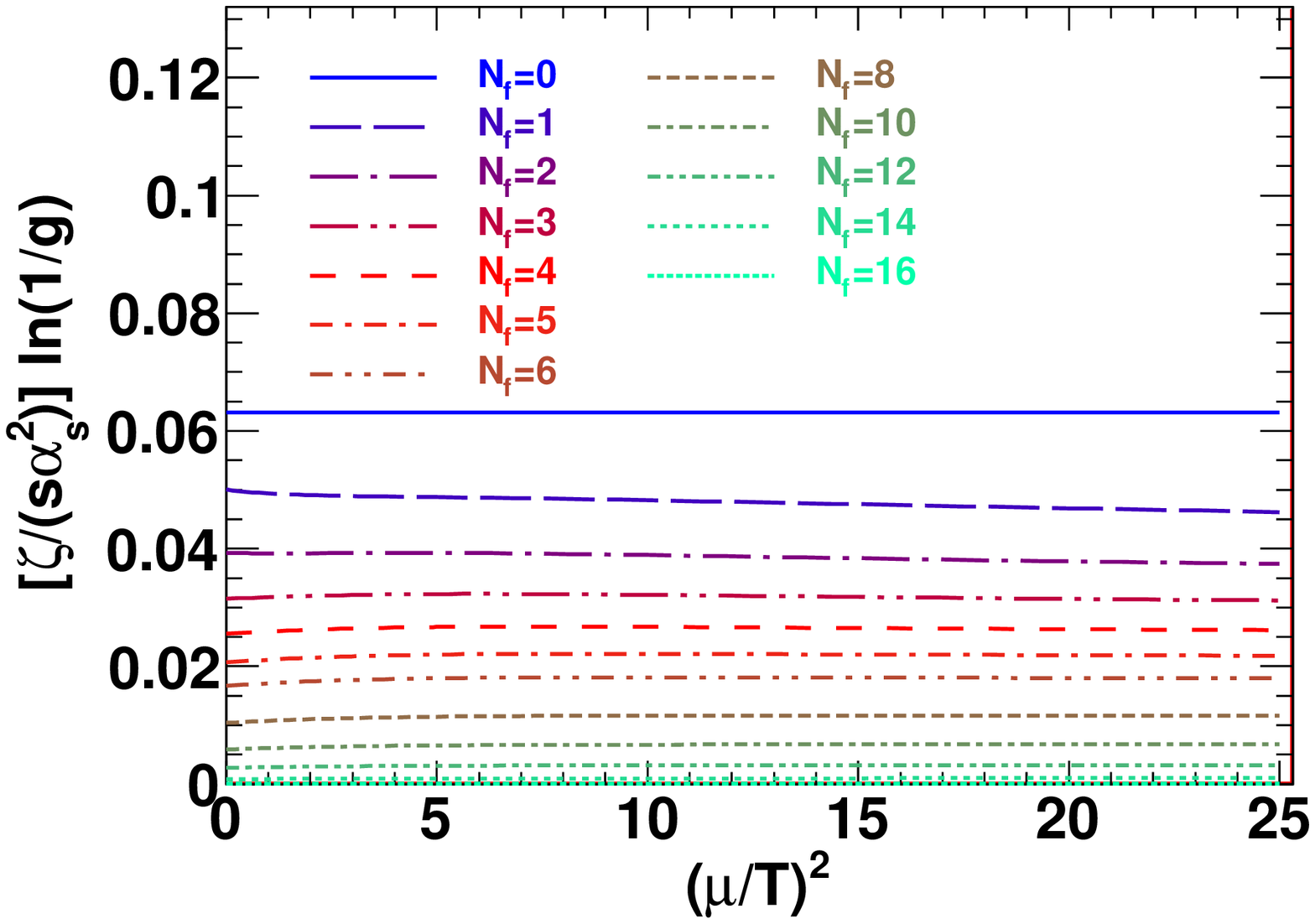}
\caption{(color online) $(\zeta/s\alpha_{s}^{2})\ln g^{-1}$ as
functions of $(\mu/T)^{2}$ for different $N_{f}$.}
\label{bulk over s with musquare}
\end{center}
\end{figure}

In Fig.\ \ref{bulk viscosity with musquare} we show the normalized bulk
viscosity $\tilde{\zeta}\equiv (\zeta /\alpha_{s}^{2}T^{3})\ln g^{-1} $ as
functions of $(\mu/T)^{2}$ for different $N_{f}$. When $\mu/T\rightarrow0$, $\zeta$ scales as $T^{3}$ from
dimensional analysis. When $T/\mu\rightarrow0$, $\zeta$ scales as $\mu^{2}T$. Technically, this is because
as $T\rightarrow 0$, $f^{g}=f^{\bar{q}}=0$, the dominant contribution in Eq.(%
\ref{bulk''}) comes from the scattering between quarks. In quark scattering,
the combination $K \equiv
f_{k_{1}}^{a}f_{k_{2}}^{b}F_{k_{3}}^{c}F_{p}^{d}\left[
A^{a}(k_{1})+A^{b}(k_{2})-A^{c}(k_{3})-A^{d}(p)\right] ^{2}=\mathcal{O}%
(T^{2})$ in Eq.(\ref{definition Dkhi ab-cd}). This is because when $%
T\rightarrow 0$,  the scattering can only happen on the fermi surface
otherwise it will be Pauli blocked (this is imposed by the vanishing of the
prefactor $f_{k_{1}}^{a}f_{k_{2}}^{b}F_{k_{3}}^{c}F_{p}^{d}$). But
scattering on the fermi surface yields $%
A^{a}(k_{1})=A^{b}(k_{2})=A^{c}(k_{3})=A^{d}(p)$ and thus $K=0$. Therefore,
the dimensionless combination $K$ contributes at
$\mathcal{O}(T^{2}/\mu ^{2})$ in Eq.(\ref{definition Dkhi ab-cd}), which leads to $\zeta \propto
T$. (A
similar argument can be applied to $\eta $. As $T\rightarrow 0$, $K^{\prime
}\equiv f_{k_{1}}^{a}f_{k_{2}}^{b}F_{k_{3}}^{c}F_{p}^{d}\left[
B_{ij}^{a}(k_{1})+B_{ij}^{b}(k_{2})-B_{ij}^{c}(k_{3})-B_{ij}^{d}(p)\right]
^{2}=\mathcal{O}(T^{0})$ in Eq.(\ref{definition Deta ab-cd}), since the
scattering on the fermi surface does not have to be forward scattering thus $%
K^{\prime }$ does not have to vanish. This leads to $\eta \propto 1/T$ in
Eq.(\ref{definition Deta ab-cd}).)
In contrast to
$\eta$, $\zeta$ is not monotonically increasing with $N_{f}$ because $\zeta$
is suppressed by an additional power of $(T_{\mu}^{\mu})^{2}
\propto\beta^{2}(g^{2})\propto (33-2N_{f} )^{2}$.
Thus when $N_{f}$ is small, $\zeta$ increases with $N_{f}$ because
quarks tend to make the averaged effective coupling weaker but at large
$N_{f}$, the suppression factor $\beta^{2}\left(  g^{2}\right)  $ takes
control to make $\zeta$ decrease with $N_{f}$. The maximum $\zeta$ happens
when $N_{f}=5$ or $6$, depending on the value of $\mu/T$.

Around $\mu=0$, we make a Taylor expansion $\tilde{\zeta}=a_{\zeta}+b_{\zeta}(
\mu/T)^{2}+\cdots$, where $a_{\zeta}$ is $\tilde{\zeta}$ at zero quark chemical potential.
The values of $a_{\zeta}$ and $b_{\zeta}$ for varies
$N_{f}$ is tabulated in Table \ref{power expansion coefficients bulk}.
Our $a_{\zeta}$ is identical to ADM's at least to the second decimal place for all
$N_{f}$ computed in Ref.\ \cite{Arnold:2006fz}. The agreement is better than $1\%$.

In Fig. \ref{bulk over s with musquare}, if the coupling $g$ is fixed,
our $\zeta/s$ is monotonically decreasing with $N_{f}$. This pattern is
qualitatively consistent with the massless pion gas result of
Ref.\ \cite{Chen:2007kx} which has
$\zeta/s\propto T^{4}/N_{f}^{2}f_{\pi}^{4}\propto 1/N_{c}^{2}N_{f}^{2}$
(only valid for $N_{f}\geq2$ where pions exist).

\section{Conclusion}

We have calculated the shear and bulk viscosities of a weakly
coupled quark gluon plasma at the leading-log order with finite temperature
$T$ and quark chemical potential $\mu$. We have found that when normalized by the
entropy density $s$, $\eta/s$ increases monotonically with $\mu$ and
eventually scales as $\left(  \mu/T\right)  ^{2}$ at large $\mu$.
However $\zeta/s$ is insensitive to $\mu$. Both $\eta/s$ and $\zeta/s$ are
monotonically decreasing function of the quark flavor number $N_{f}$ when
$N_{f}\geq2$. The same property is also observed in pion gas calculations.
Our perturbative calculation suggests that QCD becomes the most perfect (with
the smallest $\eta/s$) at $\mu=0$ and $N_{f}=16$ (the maximum $N_{f}$ with
asymptotic freedom). It would be interesting to test whether the currently
smallest $\eta/s$ computed close to the phase transition at $\mu=0$ and
$N_{f}=0$ can be further reduced by increasing $N_{f}$.

Acknowledgement: JWC and YFL thank the INT, Seattle, for hospitality. JWC and YFL
are supported by the NSC, NCTS, and CASTS of ROC. YKS is supported in part by
the CCNU-QLPL Innovation Fund (QLPL2011P01). QW is supported in part by the
National Natural Science Foundation of China under grant No. 11125524.

\newpage

\appendix

\section{Scattering Amplitudes and Taylor Expansion Coefficients of
Viscosities}

\renewcommand{\arraystretch}{1.25}
\begin{table}[ptb]
\caption{Matrix elements squared for two particle scattering processes
in QCD. The helicities and colors of all initial and final state particles are summed over.
$q_{1}$ and $q_{2}$ represent quarks of distinct flavors, $\bar{q}_{1}$ and
$\bar{q}_{2}$ are the associated antiquarks, and $g$ represents a gluon.
$d_{F}$ and $d_{A}$ denote the dimensions of the fundamental and adjoint
representations of $SU_c(N)$ gauge group while $C_{F}$ and $C_{A}$
are the corresponding quadratic Casimirs. In a $SU_c(3)$ theory
with fundamental representation fermions, $d_{F}=C_{A}=3$, $C_{F}=4/3$, and $d_{A}=8$.}
\label{amp}
\centering{}%
\begin{tabular}
[c]{|l|l|}\hline
$ab\rightarrow cd$ & $\left\vert M_{a\left(  k_{1}\right)  b\left(
k_{2}\right)   \rightarrow c\left(  k_{3}\right)  d\left(
k_{4}\right)  }\right\vert ^{2}$\\\hline
$%
\begin{array}[c]{c}
q_{1 }q_{2}\rightarrow q_{1} q_{2}\\
\bar{q}_{1 }q_{2}\rightarrow\bar{q}_{1} q_{2}\\
q_{1 }\bar{q}_{2}\rightarrow q_{1} \bar{q}_{2}\\
\bar{q}_{1 }\bar{q}_{2}\rightarrow\bar{q}_{1} \bar{q}_{2}%
\end{array}
$ & $8g^{4}\dfrac{d_{F}^{2}C_{F}^{2}}{d_{A}}\left(  \dfrac{s^{2}+u^{2}}{t^{2}%
}\right)  $\\\hline
$%
\begin{array}[c]{c}
q_{1 }q_{1}\rightarrow q_{1}q_{1}\\
\bar{q}_{1 }\bar{q}_{1}\rightarrow\bar{q}_{1}\bar{q}_{1}%
\end{array}
$ & $8g^{4}\dfrac{d_{F}^{2}C_{F}^{2}}{d_{A}}\left(  \dfrac{s^{2}+u^{2}}{t^{2}%
}+\dfrac{s^{2}+t^{2}}{u^{2}}\right)  +16g^{4}d_{F}C_{F}\left(  C_{F}%
-C_{A}/2\right)  \dfrac{s^{2}}{tu}$\\\hline
$q_{1}\bar{q}_{1}\rightarrow q_{1}\bar{q}_{1}$ & $8g^{4}\dfrac{d_{F}^{2}%
C_{F}^{2}}{d_{A}}\left(  \dfrac{s^{2}+u^{2}}{t^{2}}+\dfrac{t^{2}+u^{2}}{s^{2}%
}\right)  +16g^{4}d_{F}C_{F}\left(  C_{F}-C_{A}/2\right)  \dfrac{u^{2}}{st}%
$\\\hline
$q_{1 }\bar{q}_{1}\rightarrow q_{2}\bar{q}_{2}$ & $8g^{4}\dfrac
{d_{F}^{2}C_{F}^{2}}{d_{A}}\dfrac{t^{2}+u^{2}}{s^{2}}$\\\hline
$q_{1 }\bar{q}_{1}\rightarrow g g$ & $8g^{4}d_{F}C_{F}%
^{2}\left(  \dfrac{u}{t}+\dfrac{t}{u}\right)  -8g^{4}d_{F}C_{F}C_{A}\left(
\dfrac{t^{2}+u^{2}}{s^{2}}\right)  $\\\hline
$%
\begin{array}[c]{c}
q_{1 }g\rightarrow q_{1}g\\
\bar{q}_{1 }g\rightarrow\bar{q}_{1}g
\end{array}
$ & $-8g^{4}d_{F}C_{F}^{2}\left(  \dfrac{u}{s}+\dfrac{s}{u}\right)
+8g^{4}d_{F}C_{F}C_{A}\left(  \dfrac{s^{2}+u^{2}}{t^{2}}\right)  $\\\hline
$gg \rightarrow gg$ & $16g^{4}d_{A}C_{A}^{2}\left(  3-\dfrac{su}{t^{2}%
}-\dfrac{st}{u^{2}}-\dfrac{tu}{s^{2}}\right)  $\\\hline
\end{tabular}
\ \end{table}
\renewcommand{\arraystretch}{1.}

\begin{table}[ptb]
\caption{First two coefficients in the Taylor expansion $\tilde{\eta}=a_{\eta}
+b_{\eta}(\mu/T)^{2}+\cdots$ near $\mu=0$. Our result is identical to
AMY's \cite{Arnold:2000dr} to at least the second decimal place. }
\centering{}
\begin{tabular}
[c]{|l|l|l||l|l|l|}\hline
$N_{f}$ & $a_{\eta}$ & $b_{\eta}$ & $N_{f}$ & $a_{\eta}$ & $b_{\eta}$\\\hline
 0 & 27.125 &   0    &  9   & 172.564 &  50.381 \\\hline
 1 & 60.808 &  16.619 & 10 & 178.839 &  51.301 \\\hline
 2 & 86.472 &  27.281 & 11 & 184.389 &  52.028 \\\hline
 3 & 106.664 &  34.454 & 12 & 189.333 &  52.608 \\\hline
 4 & 122.957 &  39.459 & 13 & 193.764 &  53.074 \\\hline
 5 & 136.380 &  43.055 & 14 & 197.760 &  53.450 \\\hline
 6 & 147.627 &  45.703 & 15 & 201.380 &  53.755 \\\hline
 7 & 157.187 &  47.690 & 16 & 204.675 &  54.003 \\\hline
 8 & 165.412 &  49.207 & ~  & ~ & ~  \\\hline
\end{tabular}
\label{power expansion coefficients eta}
\end{table}

\begin{table}[ptbptb]
\caption{power expansion coefficients of $\tilde{\zeta}$ where
$\tilde{\zeta}=a_{\zeta}+b_{\zeta}(\mu/T)^{2}+\cdots$. Our result is identical to
ADM's result \cite{Arnold:2006fz} to at least the second decimal place. }
\centering{}
\begin{tabular}
[c]{|l|l|l||l|l|l|}\hline
$N_{f}$ & $a_{\zeta}$ & $b_{\zeta}$ & $N_{f}$ & $a_{\zeta}$ & $b_{\zeta}$\\\hline
  0 &   0.4430 &   0 &  9 &   0.3847 &   0.0873\\\hline
  1 &   0.5816 &   0.0393 & 10 &   0.3113 &   0.0725\\\hline
  2 &   0.6379 &   0.0747 & 11 &   0.2389 &   0.0569\\\hline
  3 &   0.6568 &   0.0981 & 12 &   0.1706 &   0.0414\\\hline
  4 &   0.6495 &   0.1116 & 13 &   0.1096 &   0.0270\\\hline
  5 &   0.6218 &   0.1172 & 14 &   0.0592 &   0.0148\\\hline
  6 &   0.5778 &   0.1163 & 15 &   0.0225 &   0.0057\\\hline
  7 &   0.5213 &   0.1103 & 16 &   0.0026 &   0.0007\\\hline
  8 &   0.4558 &   0.1003 & ~ &   ~ &   ~\\\hline
\end{tabular}
\label{power expansion coefficients bulk}
\end{table}

To describe the microscope scattering processes in a quark gluon plasma,
we need scattering amplitudes between quarks and
gluons. In a hot QCD plasma the infrared singularity in the amplitude can be
regularized by hard thermal loop dressed propagator.
We use the same amplitude (shown in Table \ref{amp})
as Ref.\ \cite{Arnold:2003zc}.

Around $\mu=0$, we make Taylor expansions $\tilde{\eta}\equiv (\eta/T^{3})
g^{4}\ln g^{-1}=a_{\eta}+b_{\eta}\left(  \mu/T\right)^{2}+\cdots $ and
$\tilde{\zeta}\equiv (\zeta/\alpha_{s}^{2}T^{3})\ln g^{-1} =a_{\zeta}+b_{\zeta}\left(  \mu/T\right)
^{2}+\cdots$.  The values of $a_{\eta}$, $b_{\eta}$, $a_{\zeta}$ and $b_{\zeta}$ for various $N_{f}$
are tabulated in Table \ref{power expansion coefficients eta} and \ref{power expansion coefficients bulk}.
Our $a_{\eta}$ and $a_{\zeta}$ are  identical to AMY's and ADM's to at
least the second decimal place for all $N_{f}$ computed in
Ref.\ \cite{Arnold:2000dr} and \cite{Arnold:2006fz}, respectively. The agreement is better than $1\%$.

\end{document}